\documentclass[apj]{emulateapj}
\usepackage{color}
\usepackage{epsfig}

\renewcommand{\thefootnote}{\fnsymbol{footnote}}

\shorttitle{Particle Acceleration at Parallel Shocks}
\shortauthors{Guo and Giacalone}

\doublespace
\begin{document}

\title{The Acceleration of Thermal Protons at Parallel Collisionless Shocks: Three-dimensional Hybrid Simulations}

\author{Fan Guo \altaffilmark{1,2} and Joe Giacalone \altaffilmark{2}}

\altaffiltext{1}{Theoretical Division, Los Alamos National Laboratory, Los Alamos, NM 87545}

\altaffiltext{2}{Department of Planetary Sciences and Lunar and Planetary
Laboratory, University of Arizona, 1629 E. University Blvd., Tucson, AZ 85721}

\email{guofan.ustc@gmail.com}

\begin{abstract}
We present three-dimensional hybrid simulations of collisionless shocks that propagate parallel to the background magnetic field to study the acceleration of protons that forms a high-energy tail on the distribution. We focus on the initial acceleration of thermal protons and compare it with results from one-dimensional simulations. We find that for both one- and three-dimensional simulations, particles that end up in the high-energy tail of the distribution later in the simulation gained their initial energy right at the shock. This confirms previous results but is the first to demonstrate this using fully three-dimensional fields. The result is not consistent with the ``thermal leakage" model. We also show that the gyrocenters of protons in the three-dimensional simulation can drift away from the magnetic field lines on which they started due to the removal of ignorable coordinates that exist in one- and two-dimensional simulations. Our study clarifies the injection problem for diffusive shock acceleration. 
\end{abstract}


\section{Introduction}

Collisionless shocks and the associated acceleration of charged particles are among the most important processes in space physics and astrophysics \citep{Blandford1987}. The theory of diffusive shock acceleration \citep[DSA;][]{Krymsky1977,Axford1977,Bell1978,Blandford1978} describes quantitatively the acceleration process in the vicinity of shock waves. It predicts a power-law energy spectrum downstream of the shock, which is a common characteristic of the energetic charged particles observed in space. However, DSA does not address the physical processes involved with the acceleration of thermal and/or low energy particles which, presumably, constitute the source of the high-energy particles. This is known as the ``injection problem'' which has received a lot of recent attention but, as yet, there is not a common consensus on its resolution \citep[e.g.,][]{Kirk2001}.

Collisionless shocks propagating in magnetized plasmas are usually divided into two classes based on the angle, $\theta_{Bn}$, between the incident magnetic field vector and shock normal vector. Quasi-parallel shocks are those for which $0^\circ \le \theta_{Bn} < 45^\circ$ and quasi-perpendicular shocks are for $45^\circ < \theta_{Bn} \le 90^\circ$. For quasi-parallel shocks, there have been several mechanisms proposed in order to solve the injection problem. It is usually thought that for parallel shocks Alfven waves excited by the streaming of protons can scatter ions in pitch angle. The particles can be accelerated by DSA when their anisotropy is small enough. \citet{Ellison1981} first proposed a Monte Carlo model for DSA that includes the injection process, where the injected particles originate from the shock-heated ions downstream of the shock that then leak back upstream of the shock. Subsequently, the particles are ``injected'' into the standard DSA process of scattering back and forth across the shock. The model assumes that all the particles are scattered by magnetic fluctuations and ignores the details of the shock structure. It has been used to fit and compare with observations at Earth's bow shock \citep{Ellison1985a,Ellison1990}. Similar models have also been discussed by other authors \citep[e.g.,][]{Malkov1998,Kang2012}. This is usually referred to as the ``thermal leakage" model. 

However, a number of authors have found a different scenario for the initial energization at parallel shocks based on self-consistent hybrid simulations  \citep{Quest1988,Scholer1990a,Scholer1990b,Kucharek1991,Giacalone1992,Sugiyama1999,Su2012}. They find that the accelerated ions originate from the shock transition layer rather than the downstream region. Some initially thermal ions are accelerated to high energies as they execute cycloidal motion within the electric and magnetic fields in the shock layer. Although the average incident magnetic field is parallel to the shock normal, as the enhanced upstream magnetic fluctuations steepen and convect through the shock layer, the angle between the incident magnetic field and the shock normal right at the shock front can significantly deviate from the average value and even be close to a quasi-perpendicular shock locally \citep[e.g.,][]{Wilson2013}. A particle can gain the first amount of energy by moving against the electric field within the shock layer \citep{Kucharek1991,Giacalone1992}. It has been clearly shown by \citet{Kucharek1991} that most of the accelerated particles are reflected by the shock and gain the first increment of energy at the shock layer itself. It is noteworthy that \citet{Lyu1990} also used hybrid simulations and found that the leakage of protons dominated the accelerated particles. We suggest that the reason they obtained a different result from other authors may be due to the method they used to drive shocks. In their simulations, the upstream and downstream plasmas are initially connected smoothly by assuming a hyperbolic tangent function for the shock parameters with a thickness of several ion inertial lengths. The magnetic fluctuations that are important to reflect ions at shock front are ignored at the beginning of the simulation. Since the simulation only lasts for $85 \Omega_{ci}^{-1}$, where $\Omega_{ci}$ is the gyrofrequency of protons in the asymptotic upstream magnetic field, it is difficult to determine if the leakage of protons can dominate the accelerated population of particles after the shock is fully developed. There is observational evidence in support of the idea that particles originate in the shock layer. For example, signatures of ion reflection at quasi-parallel portion of Earth's bow shock has been reported \citep{Gosling1982,Gosling1989b}. It is worth noting that although the initial acceleration in these two processes are quite different, these two models are closely connected and it is difficult to distinguish between them in observations. The Monte Carlo model has been explicitly compared with the results of the hybrid simulations. The comparison generally has a good agreement but some obvious differences \citep{Ellison1993}. The readers are referred to \citep{Burgess2012} for a complete review of these two models.

The injection process at quasi-perpendicular shocks is thought to be more complicated. The required pre-acceleration may be achieved by processes such as shock drift acceleration \citep[e.g.,][]{Armstrong1985,Decker1988} and shock surfing acceleration \citep{Lee1996,Zank1996}. In shock drift acceleration, charged particles drift because of the gradient in the magnetic field at the shock front. The direction of the drift is in the same direction as the motional electric field vector $\textbf{E} = - \textbf{V} \times \textbf{B} /c$, and the particles gain energy during this drift motion. In shock surfing acceleration, it is thought that the cross shock potential electric field is a barrier to some particles incident on the shock leading to their reflection at the shock and a drift against the motional electric field leading to the particle acceleration. Recent progress has been made to distinguish the relative importance for these two processes. For instance, it has been found that in order for shock surfing acceleration to be efficient, the thickness of the shock layer has to be very thin -- on the scale of an electron inertial length \citep{Lipatov1999}, which is not consistent with the shocks observed in space and in numerical simulations \citep{Bale2003,Leroy1982}. Moreover, it has been shown that the shock thickness has to be fairly large compared to electron gyroradii to be consistent with the observation of electron heating at shocks \citep{Lembege2004}. Using one-dimensional hybrid simulations, \citet{Wu2009} showed that the reflection by cross shock potential electric field is inefficient for suprathermal particles. \citet{Yang2009} and \citet{Yang2012} have used one-dimensional and two-dimensional full particle simulations with test-particle simulations to show that shock drift acceleration dominates the initial acceleration and shock surfing acceleration can sometimes have a considerable contribution.

\citet{Giacalone1999} have demonstrated that large-scale magnetic turbulence which is present in the background plasma can efficiently lower the injection threshold at perpendicular shocks by increasing the transport of charged particles normal to magnetic field. Recent numerical simulations for the acceleration of charged particles (both ions and electrons) in the existence of large-scale magnetic fluctuations show very efficient acceleration, which indicates that there is \textit{no} injection problem at perpendicular shocks \citep{Giacalone2005a,Giacalone2005b,Guo2010a,Guo2012b,Guo2012c}. It is worth noting that observationally there seems \textit{no} injection problem at quasi-perpendicular shocks. For instance, the Voyager spacecraft clearly show energetic particles (at least several $MeV$) are accelerated at the solar wind termination shock \citep{Decker2005,Decker2008}. Strong interplanetary shocks observed by ACE and Wind at $1$ AU (mostly quasi-perpendicular shocks) can accelerate protons to energies more than $50$ keV \citep{Giacalone2012b}.

It is important to point out that, in order to accurately model the acceleration of charged particles at shocks, one has to consider fully three-dimensional electromagnetic fields. This is because in a field that has at least one ignorable coordinate, the motions of particles are artificially restricted in the sense that any individual charged particles must remain within one gyroradius of the magnetic field line on which it began its motion \citep{Jokipii1993,Giacalone1994a,Jones1998}. Many previous hybrid simulations were performed using only one or two spatial dimensions (Scholer et al., 1993), and, thus, are subject to this constraint. The effect of cross-field diffusion must be included and requires fully three-dimensional simulations \citep{Giacalone2000b}. Recent two-dimensional and three-dimensional particle-in-cell simulations have shown signatures of energetic particles at parallel shocks but did not examine the initial acceleration process at the shock front \citep{Niemiec2012,Caprioli2013}.

In this study, we employ a three-dimensional self-consistent hybrid simulation to study the initial particle energization at a parallel shock. We find that all particles accelerated to high energies gain the first amount of energy right at the shock front. In the three-dimensional simulation, the accelerated protons can move off their original field lines. Nevertheless, the initial acceleration process is consistent with that is found in previous one-dimensional and two-dimensional hybrid simulations. This finding clarifies the acceleration of low energy particles and the injection process for DSA found in other works \citep[e.g.,][]{Caprioli2013}.

\section{Numerical Method}

\begin{figure}
\centering
\begin{tabular}{c}
\epsfig{file=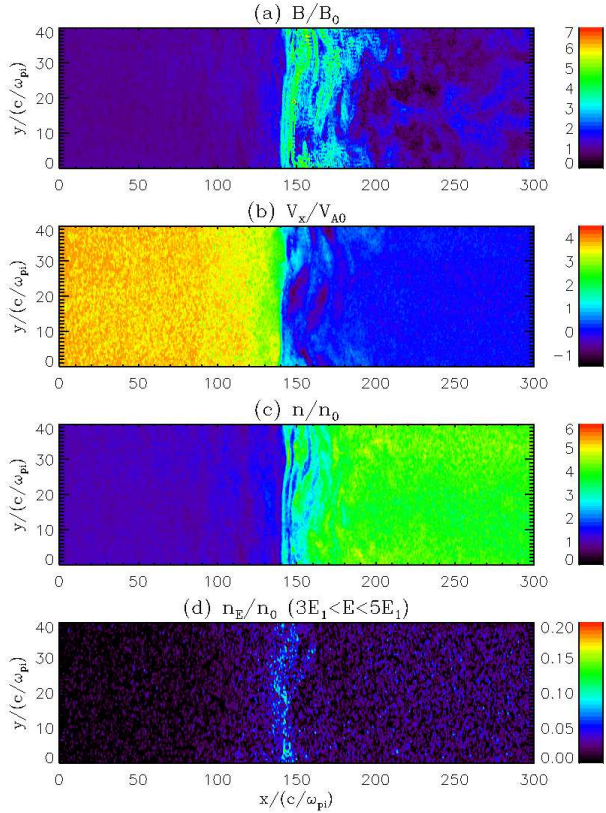,width=80mm,clip=}
\end{tabular}
\caption{Two-dimensional representation of the three-dimensional hybrid simulation in the $z = 20$ $c/\omega_{pi}$ plane at $\Omega_{ci}t = 120.0$: (a) the magnitude of the magnetic field $B/B_0$, (b) the $x$-component of the bulk proton velocity $V_x/V_{A0}$, (c) the density of the plasma flow $n/n_0$, and (d) the density of the accelerated particles with energy $3 E_1 < E < 5 E_1$, where $E_{1}$ represents the upstream proton ram energy $m_p U_1^2 /2$ in the shock frame, and $U_1$ is taken to be $5.3 V_{A0}$. \label{Fig1}}
\end{figure}

We perform three-dimensional hybrid simulations of parallel shocks and compare the results with that of one-dimensional hybrid simulations. Although our simulations resolve the full proton distribution function, we are particularly interested in those particles that are accelerated to high energies. In the hybrid simulations \citep[e.g.,][]{Winske1988}, the ions are treated kinetically and thermal electrons are treated as a massless fluid. This approach is well suited to resolve ion-scale plasma physics, which is crucial to describe supercritical collisionless shocks and particle acceleration from thermal to suprathermal energies. We have improved the efficiency of our one-dimensional, two-dimensional, and three-dimensional hybrid simulation models \citep[Section 3.3.1 in][]{Guo2012}. The new codes have been implemented and tested on the NASA's Pleiades supercomputer using a few thousand CPU cores. For the three-dimensional simulations, we consider a three-dimensional Cartesian grid ($x, y, z$). All the physical vectors such as the positions and velocities of protons, and electric and magnetic fields $\textbf{E}$ and $\textbf{B}$ have components in three directions and also spatially depend on $x, y,$ and $z$. For the one-dimensional simulations, the physical vectors have components in three directions but only depend on $x$. A shock is produced by using the so-called piston method, in which the plasma is injected continuously from one end of the simulation box ($x=0$, in our case), and reflected elastically at the other end ($x=L_x$). The right boundary is assumed to be a perfectly conducting barrier. The pileup of density and magnetic field creates a shock propagating in the $-x$ direction. In the three-dimensional simulation, the boundary conditions of the electromagnetic fields in the $y$ and $z$ direction are periodic, and the particles that move out of one end of the simulation domain in $y$ or $z$ direction will re-enter the domain from the other end. 

We examine two different simulation cases (one-dimensional and three-dimensional) with similar parameters. In both cases, the Mach number of the flow in the simulation frame is $M_{A0} = V_x/V_{A0} = 4.0$, where $V_{A0}$ is the upstream Alfven speed. The electron and ion plasma betas are $\beta_e = 1.0$ and $\beta_i = 0.5$, respectively. The grid size is $\Delta x \times \Delta y \times \Delta z = 0.5 c/\omega_{pi} \times 0.5 c/\omega_{pi} \times 0.5 c/\omega_{pi}$ for the three-dimensional case and $\Delta x = 0.5 c/\omega_{pi}$ for the one-dimensional simulation case, where $c/\omega_{pi}$ is the ion inertial length, and $c$ and $\omega_{pi}$ are the light speed and proton plasma frequency, respectively. The time step is taken to be $\Omega_{ci} \Delta t = 0.01$, where $\Omega_{ci}$ is the proton gyro-frequency. The ratio between light speed and upstream Alfven speed is $c/v_{A0} = 6000.0$, and the anomalous resistivity is $\eta = 1\times 10^{-6} 4\pi \omega_{pi}^{-1}$. The initial spatially uniform thermal ion distribution is represented using $25$ particles per cell in the three-dimensional case and $200$ particles per cell in the one-dimensional case. We have repeated the simulations with different number of particles per cell and found it does not change the results. Initially the average magnetic field is assumed to be $\textbf{B}_0 = B_0 \hat{x}$, i.e., the average shock-normal angle is zero.  The spatial size in the $x$ direction is taken to be $300 c/\omega_{pi}$ for both of the cases. For the three-dimensional case, the lengths of the simulation box in the $y$ and $z$ directions are taken to be $L_y = L_z = 40 c/\omega_{pi}$. Pre-existing magnetic fluctuations are not included \citep{Giacalone1992}.

 \begin{figure}
 \begin{center}
\includegraphics[width=80mm]{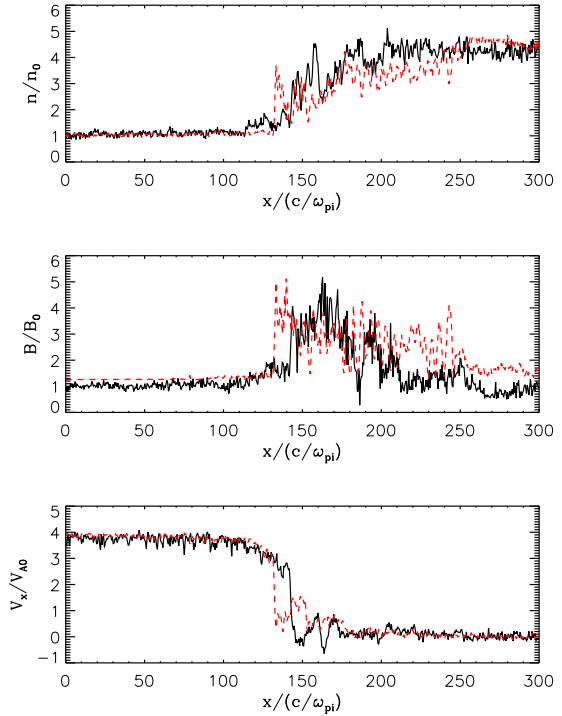}
\caption{One-dimensional $x$-profiles from the three-dimensional simulation at $y = 20 c/\omega_{pi}$ and $z = 20 c/\omega_{pi}$ (black solid lines) and the one-dimensional simulation (red dashed lines): (a) plasma number density $n/n_0$, (b) the magnitude of the magnetic field $B/B_0$, and (c) the $x$-component of the bulk proton velocity $V_x/V_{A0}$. \label{Fig2}}
 \end{center}
 \end{figure}

 \begin{figure}
\centering
\begin{tabular}{c}
\epsfig{file=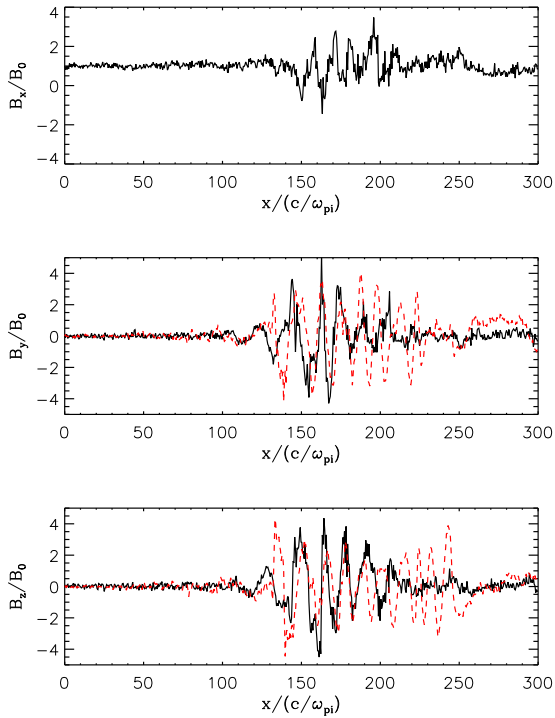,width=80mm,clip=}
\end{tabular}
\caption{Similar to Figure \ref{Fig2}, but for $x$-profiles of (a) the $x$-component of the magnetic field $B_x/B_0$, (b) the $y$-component, $B_y/B_0$, and (c) the $z$ component, $B_z/B_0$. Note in one-dimensional simulations, $B_x$ is always taken to be $B_0$ (not shown). \label{Fig3}}
\end{figure}

\section{Simulation Results}

Figure \ref{Fig1} shows results from the three-dimensional simulation. Shown are representations of various quantities in the x-y plane at $z = 20$ $c/\omega_{pi}$ at $\Omega_{ci}t = 120.0$. The figure shows the color-coded contours of (a) the magnitude of the magnetic field $B/B_0$, (b) the x-component of the bulk proton velocity $V_x/V_{A0}$, (c) the plasma density $n/n_0$ and (d) the density of accelerated particles $n_E/n_0$ with energies of $3 E_1 < E < 5 E_1$, where $E_{1}$ represents the upstream proton ram energy $m_p U_1^2 /2$ in the shock frame, and $U_1$ is taken to be $5.3 V_{A0}$, which is the upstream plasma flow speed measured in the frame of the shock. From these plots we can see that the shock front is approximately located between $x = 140$ - $150 c/\omega_{pi}$ where the ion density and flow speed have clear, abrupt transitions.  The magnetic field and fluid velocity in the shock region show strong fluctuations. The fluctuations also modify the spatial distribution of plasma density and the density of accelerated particles. The accelerated protons are concentrated around the shock layer where the magnetic field increases, indicating a strong localized energization in this region.

\begin{figure}
\centering
\begin{tabular}{c}
\epsfig{file=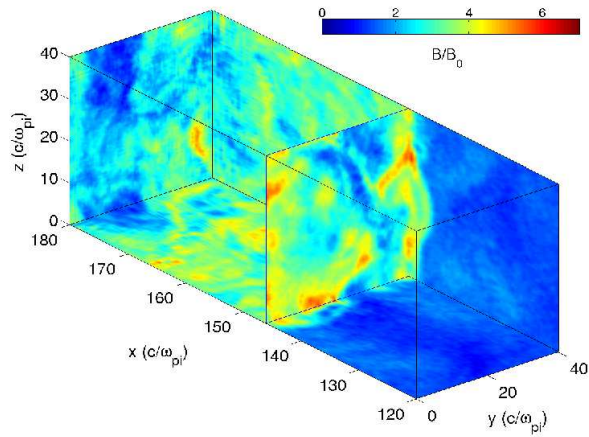,width=80mm,clip=}
\end{tabular}
\caption{Three-dimensional representation of the magnitude of magnetic field $B/B_0$ at $\Omega_{ci}t = 120.0$. The two-dimensional slice at $x = 146 c/\omega_{pi}$ plane shows the transition of the magnetic field. \label{Fig4-B}}
\end{figure} 

In Figure \ref{Fig2} and Figure \ref{Fig3} we compare the $x$-profiles of the three-dimensional hybrid simulation with that of the one-dimensional hybrid simulation. The black solid lines in Figure \ref{Fig2} show (a) the plasma density $n/n_0$, (b) the magnitude of magnetic field $B/B_0$, and (c) flow velocity in the $x$ direction $V_x/V_{A0}$ along $y = 20$ $c/\omega_{ci}$ and $z = 20$ $c/\omega_{ci}$ at $\Omega_{ci}t = 120.0$. The red dashed lines show the $x$-profiles from the one-dimensional simulation at $\Omega_{ci}t = 110.0$. The black solid lines in Figure \ref{Fig3} show the $x$-profiles from the three-dimensional simulation along the same line with that in Figure \ref{Fig2}, but for the magnetic field components in $B_x/B_0$, $B_y/B_0$, and $B_z/B_0$, respectively. The red dashed lines represent the results from the one-dimensional simulation at $\Omega_{ci}t = 110.0$. Note in one-dimensional simulations, $B_x$ is always taken to be $B_0$ (not shown). For the one-dimensional simulation, the location of the shock in the $x$ direction is about $x = 132 c/\omega_{ci}$ and the upstream inflow speed in the shock frame is about $U_1 = 5.5 V_{A0}$, slightly larger than that in the three-dimensional simulation. For the three-dimensional case, the magnetic field fluctuations are mostly transverse to the initial magnetic field in the $x$ direction. The fluctuating components $\delta B_y$ and $\delta B_z$ contain about $90 \%$ of the total energy of the magnetic fluctuations. The fluctuations are circularly right-handed polarized, and are convected toward the shock, indicating that they are excited by reflected ions that flow upstream. Close to the shock, the fluctuations are compressed and amplified. The downstream magnetic field consists of large-amplitude fluctuations. The results are consistent with previous analytical theories and one-dimensional numerical simulations \citep[e.g.,][]{Quest1988}. The results from three-dimensional simulations are qualitatively consistent with those from the one-dimensional simulation.
 
Figure \ref{Fig4-B} gives a three-dimensional representation of the magnitude of magnetic field $B/B_0$ close to the shock region at $\Omega_{ci}t = 120.0$. The two-dimensional slice at $x = 146 c/\omega_{pi}$ plane clearly shows the transition of the magnetic field. The shock front is highly turbulent due to its interaction with upstream fluctuations, which produces various enhancements of magnetic field at different regions. This is similar to recent results of large-scale multi-dimensional hybrid simulations \citep{Lin2005,Omidi2013,Caprioli2013}. Due to the limited box size and simulation time, the size of structures in the shock surface can only grow to $\sim 10-20 c/\omega_{pi}$, which are smaller than that in recent large-scale hybrid simulations \citep{Omidi2013,Caprioli2013}. The three-dimensional features at parallel shock regions will be further investigated in the future.
 
\begin{figure}
\centering
\begin{tabular}{c}
\epsfig{file=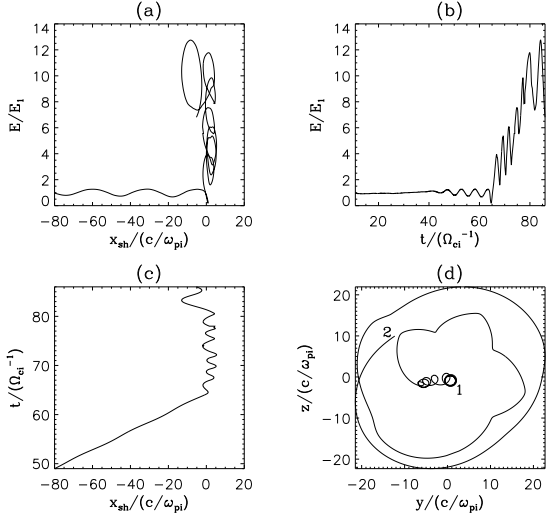,width=80mm,clip=}
\end{tabular}
\caption{The trajectory of a representative particle accelerated at the shock layer from the one-dimensional simulation. The physical quantities for each panel are: (a) the energy versus the distance away from the shock layer, (b) the energy versus the time in the simulation, (c) the simulation time versus the distance away from the shock, and (d) the location along the $z$ direction $z = \int v_z dt + z_0$ versus the location along the $y$ direction $y = \int v_y dt + y_0$, assuming the initial location is at $y_0 = 0$ and $z_0 = 0$. \label{fig5-traj1d}}
\end{figure}

\begin{figure}
\centering
\begin{tabular}{c}
\epsfig{file=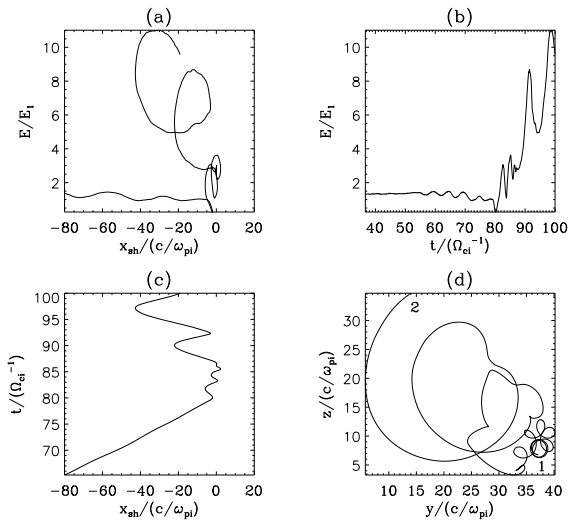,width=80mm,clip=}
\end{tabular}
\caption{The trajectory of a representative particle accelerated at the shock layer from the three-dimensional simulation. The plotted physical quantities are similar to Figure \ref{fig5-traj1d}. \label{fig6-traj3d}}
\end{figure}

Now we discuss the acceleration of protons at the shock front. \citet{Jokipii1993}, \citet{Giacalone1994a} and \citet{Jones1998} have demonstrated that in simulation models with at least one ignorable coordinate, the guiding center of an individual charged particle is confined to within one gyroradius of the magnetic field line on which it began its motion. In order to overcome this artificial constraint and to accurately model the acceleration process, it is important to consider the motions of charged particles in three-dimensional electric and magnetic fields. Since we perform three-dimensional hybrid simulations in this study, the restriction on the motions of charged particles is removed. 

\begin{figure}
\centering
\begin{tabular}{c}
\epsfig{file=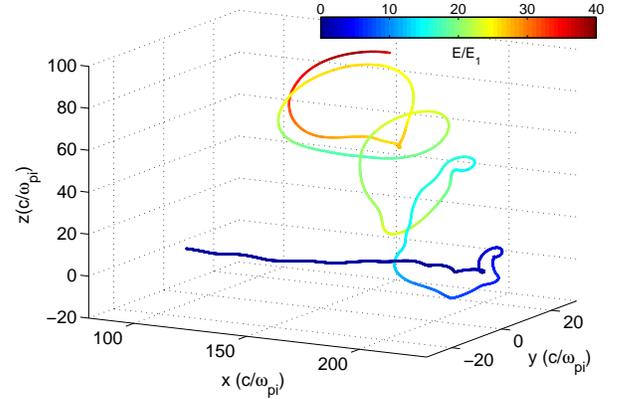,width=80mm,clip=}
\end{tabular}
\caption{Three-dimensional trajectory of an accelerated particle in the three-dimensional simulation during $\Omega_{ci}t=22.0$ -- $92.0$. The energy of the particle is represented by its color. \label{Fig7}}
\end{figure}

Figure \ref{fig5-traj1d} shows the trajectory of a representative particle in the one-dimensional simulation. The physical quantities for each panel are: (a) the energy versus the location in the $x$ direction away from the shock front in the shock frame $x_{sh}$ (the location of the shock front is assumed to be $x_{sh} = 0$), (b) the energy versus the time in the simulation, (c) the simulation time versus $x_{sh}$, and (d) the location in the $z$ direction $z = \int v_z dt + z_0$ versus the location in the $y$ direction $y = \int v_y dt + y_0$, where $y_0$ and $z_0$ are assumed to be zero, respectively. The marker `1' notes the starting point and `2' notes the end point of the trajectory. One can clearly see that the initial acceleration occurs right at the shock front $x_{sh} = 0$. The acceleration mechanism is due to the reflection at the shock, and the particle gains energy in the electric field $\textbf{E} =-\textbf{V} \times \textbf{B}/c$ since locally the angle between magnetic field vector and shock normal vector can be quite large. As shown in this figure, the particle ``rides'' along the shock front and is accelerated for about $15$ gyroperiods. The resulting energy gain is about $10$ times of the plasma ram energy $E_1$. This process has been discussed by previous authors \citep{Quest1988,Scholer1990a,Scholer1990b,Kucharek1991,Giacalone1992} and our results are consistent with the previous work. Since we use the one-dimensional hybrid simulation, the motion of the particle is restricted on its original field line and its guiding center to within one gyroradius of the initial field line.

\begin{figure}
\centering
\begin{tabular}{c}
\epsfig{file=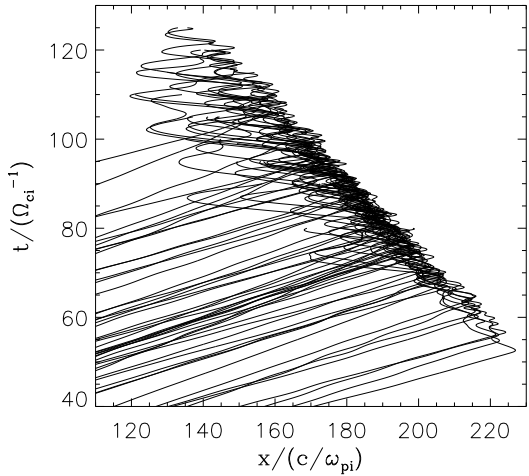,width=80mm,clip=}
\end{tabular}
\caption{The locations of $50$ accelerated particles in the x-direction as a function of simulation time. The particles 
are tracked until they gain energies several times the plasma ram energy. It shows that all the accelerated particles gain initial energies at the shock layer. \label{Fig8}}
\end{figure}

In Figure \ref{fig6-traj3d} we show the trajectory of an accelerated particle in the three-dimensional simulation. The plotted physical quantities are similar to Figure \ref{fig5-traj1d}. Since the electric and magnetic fields are fully three-dimensional, this particle (and all others) is not restricted to its original field line as in the one-dimensional simulation. This can be seen from Figure \ref{fig6-traj3d} (d), in which the guiding center of the charged particle drifts in the $y$-$z$ plane away from its original field line during the acceleration. We have examined the trajectory of this particle until the end of the simulation and found that its guiding center does not move back to within one gyroradius of the position of the field line that the particle started its gyromotion. The original field line is not likely far away (in the $y$-$z$ plane) from where it was originally. In order for this to happen, there would have to be significant flows in the $y$ and $z$ direction upstream of the shock. Figure \ref{Fig7} shows the three-dimensional trajectory of an accelerated particle in the three-dimensional simulation during $\Omega_{ci}t=22.0$ -- $92.0$. The energy of the particle is represented by its color. It clearly shows the reflection of the particle at shock front, and the particle gains energy during the gyromotion. Eventually the particle can gain a energy up to $40 E_1$ and drift several tens of ion inertial lengths away from its original location along the $z$ direction.

\begin{figure}
\centering
\begin{tabular}{cc}
\epsfig{file=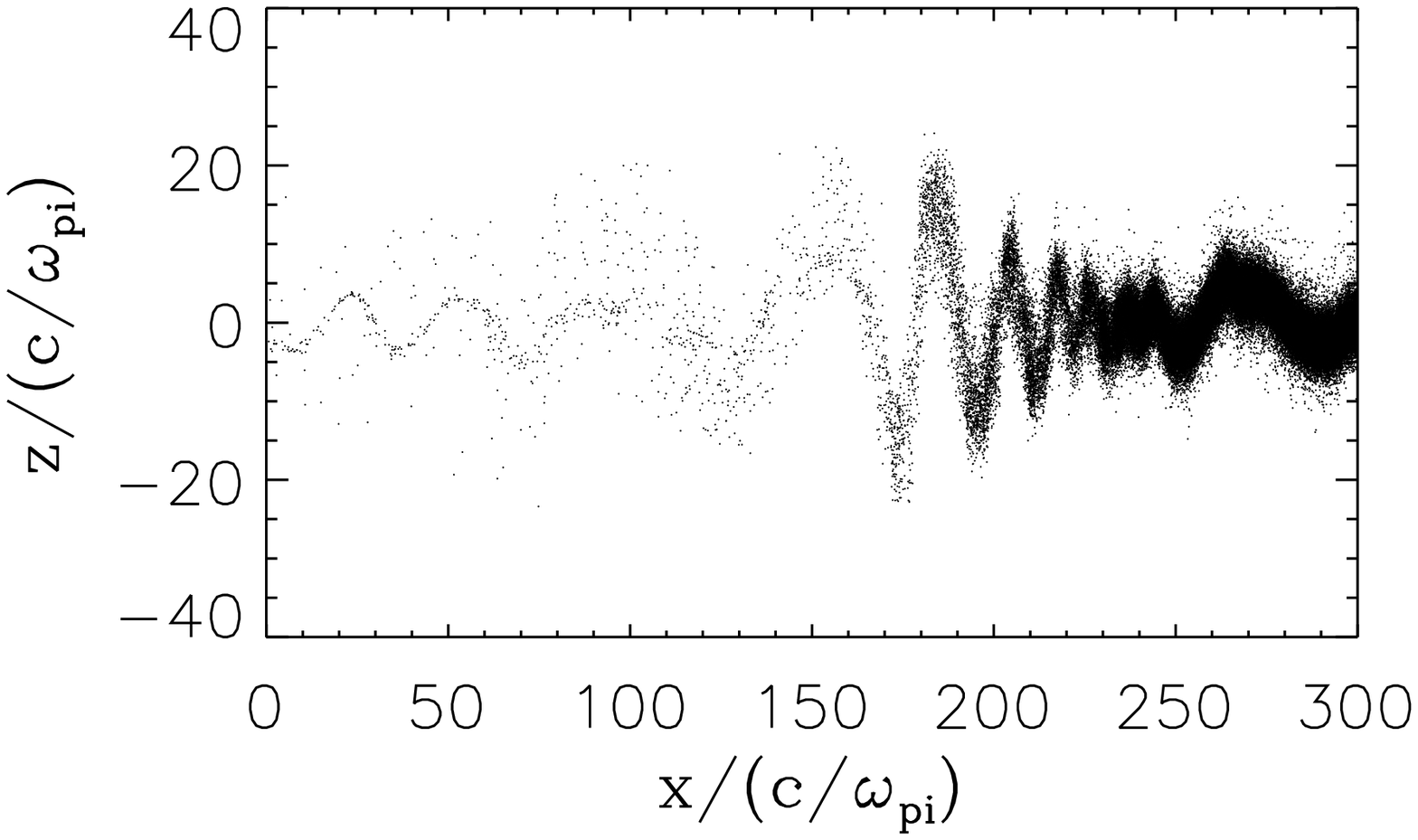,width=80mm,clip=} \\
\epsfig{file=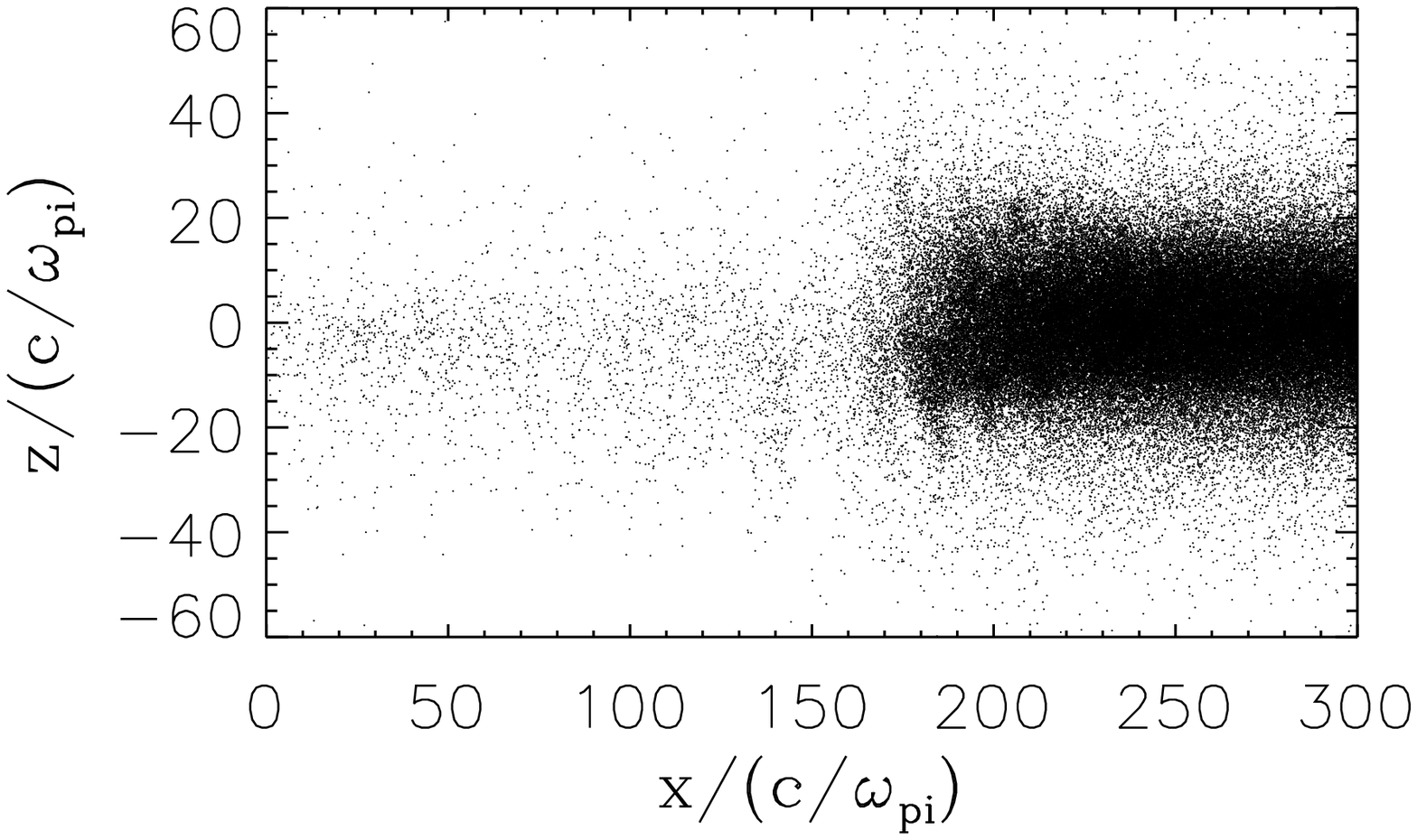,width=80mm,clip=}
\end{tabular}
\caption{The displacements of particles from their original positions in the z-direction versus the location of particles in the x-direction. The upper panel shows the results from the one-dimensional simulation. The bottom panel shows the results from the test-particles that are initially on the same field line $y=20c/\omega_{pi}$ and $z=20c/\omega_{pi}$ in the three-dimensional simulation. \label{Fig9}}
\end{figure}

However, even in the three-dimensional simulation we find that the acceleration process is quite similar to what is found in the one-dimensional simulation. The particles still gain their first amount of energy at the shock front. We have examined several hundred accelerated particles in the three-dimensional simulation and found that they all get the initial acceleration right at the shock layer. In Figure \ref{Fig8} we randomly selected $50$ accelerated particles in the three-dimensional simulation and show their positions in the $x$ direction as a function of time. The particles are tracked until they gain energies several times of the plasma ram energy. This clearly shows that the accelerated particles are originated from the shock layer, \textit{not} in the downstream region of the shock. This is consistent with previous one-dimensional hybrid simulations \citep{Kucharek1991}.

\begin{figure}
\centering
\begin{tabular}{c}
\epsfig{file=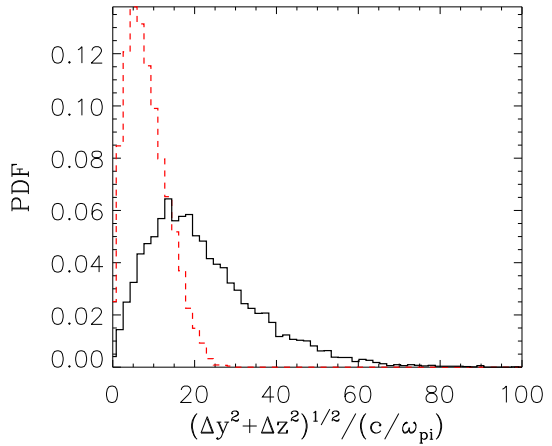,width=80mm,clip=}
\end{tabular}
\caption{The probability distribution function (PDFs) for the transverse displacements of the particles in the downstream region. The black solid line is for the test-particles originally along $y = 20c/\omega_{pi}$ and $z = 20c/\omega_{pi}$ in the three-dimensional simulation and the red dashed line is for the one-dimensional simulation. The PDFs have been normalized using the counted number of particles. \label{Fig10}}
\end{figure}

Since field lines themselves meander in space, it is difficult to demonstrate that particles move off of field line simply by examining individual trajectories. To address this, in the upper panel of Figure \ref{Fig9} we show a scatter plot for particles with energies larger than $1.5 E_1$ from the one-dimensional simulation at $\Omega_{ci}t = 100.0$ in the $x$-$z$ plane that were initially all on the same magnetic field line. This method has been used by \citet{Giacalone1994a} and \citet{Giacalone1994b} to show that in one-dimensional and two-dimensional simulations, the motions of charged particles are restricted. It can be seen that in the one-dimensional simulation, the distribution of the particles is along a fluctuating field line, meaning that they are restricted on the field line. In the three-dimensional simulation, we inject $120000$ test-particles (the same number of particles as in the one-dimensional simulation) at the beginning of the simulation and track their motions during the simulation. The test-particles are initially evenly distributed along line at $y = 20c/\omega_{pi}$ and $z = 20c/\omega_{pi}$ in the $x$ direction (along the same field line). In the bottom panel of Figure \ref{Fig9}, we show the displacements of those particles with energies larger than $1.5 E_1$ at $\Omega_{ci}t = 120.0$ similar to the one-dimensional case. We find that the distribution of the particles has no obvious structure, indicating they move off their original field line. \citet{Giacalone1994a} and \citet{Giacalone1994b} have shown that in one-dimensional and two-dimensional numerical simulations, the motions of charged particles are restricted on their original field lines. Here we show that this restriction is indeed removed in the three-dimensional hybrid simulation.  We also observe that some test-particles can move a large distance ($>40c/\omega_{pi}$) transverse to the magnetic field in the three-dimensional simulation.

\begin{figure}
\centering
\begin{tabular}{c}
\epsfig{file=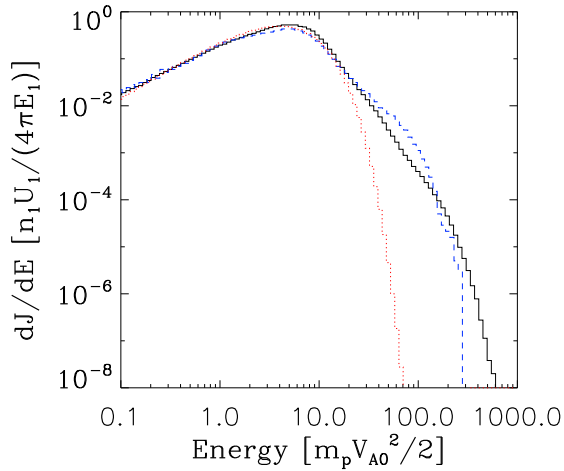,width=80mm,clip=}
\end{tabular}
\caption{Downstream energy spectra of protons in the three-dimensional simulation (black solid line) and the one-dimensional simulation (blue dashed line) at $\Omega_{ci}t = 120.0$. A Maxwellian distribution determined from shock jump relation is also shown using red dotted line. \label{Fig11-spectra}}
\end{figure}

In Figure \ref{Fig8} we plot the probability distribution functions (PDFs) for the transverse displacement from the initial location $\sqrt{\Delta y^2 + \Delta z^2}$ from the two cases shown in Figure \ref{Fig9}. The black solid line represents the results from the three-dimensional simulation and the red dashed line represents the results from the one-dimensional simulation. The PDFs have been normalized using the counted number of particles. It can be seen that a significant fraction of the particles can move a large distance ($>40c/\omega_{pi}$) in the transverse direction because the particles can move off their original field line. In the one-dimensional simulation, the transverse motion is restricted within about $20c/\omega_{pi}$. In our case the field line meandering due to the large-amplitude magnetic fluctuation can cause some particles to move transverse to the $x$ direction. However, the results for the one-dimensional simulation (Figure \ref{Fig9} and Figure \ref{Fig10}) are consistent with \citep{Giacalone1994b}, who demonstrated the particles have a restricted cross-field motion in one- and two-dimensional fields by simulating a quasi-perpendicular shock with $\theta_{Bn} = 45^\circ$.

In Figure \ref{Fig11-spectra} we show a comparison of the downstream energy spectra at $\Omega_{ci}t = 120.0$ between the one- and three-dimensional simulations. The black solid line shows the spectrum for the three-dimensional simulation and the blue dashed line represents the spectrum for the one-dimensional simulation. A Maxwellian distribution determined from shock jump relation is also shown using red dotted line. In both of the two cases, a fraction of thermal protons are accelerated to more than several times the plasma ram energy, $E_1$. The acceleration at the shock layer can have a significant contribution to nonthermal tails in the energy spectra. Since in the one-dimensional simulation we have a limited number of particles, the spectrum only extends to about $200$ $m_pV_{A0}^2/2$. For the three-dimensional simulation, the particles can be energized up to $600$ $m_pV_{A0}^2/2$. In both of the cases, the density of the accelerated particles with energy larger than $20$ $m_pV_{A0}^2$ is about $5\%$ of the downstream plasma density, indicating an efficient injection of energetic particles.

\section{Discussion and Conclusions}

We performed three-dimensional hybrid (kinetic ions and fluid electrons) simulations to investigate the initial acceleration of thermal protons at parallel shocks. We examined the trajectories of many individual particles and the spatial distribution of protons that comprise the high energy tail resulting from the acceleration of particles at the shock.  We find that the charged particles gain energy at the shock layer similar to the one-dimensional simulation. The particle can ride on the shock front and gain a large amount of energy. The results confirm previous hybrid simulations that the initial acceleration of charged particles is right at the shock front \citep{Quest1988,Scholer1990a,Scholer1990b,Kucharek1991,Giacalone1992}, even in a three-dimensional electromagnetic field. The result clarifies the injection process for diffusive shock acceleration. In addition, we find that the guiding center of accelerated particle can move off their original field lines in the three-dimensional simulation.

The ``injection problem'' is a long standing problem for diffusive shock acceleration. The charged particles are required to be energetic enough and efficiently interact with magnetic fluctuations. The results of our study bear on this problem. In this study we focus on the mechanism involved in accelerating thermal or low-energy particles at the shock front. Previous studies have proposed two different mechanisms for the injection of low-energy ions at parallel shocks, i.e., particle energization by downstream heating \citep{Ellison1981} and by reflection at the shock layer \citep{Quest1988,Scholer1990a,Scholer1990b,Kucharek1991,Giacalone1992}. Although previous hybrid simulations have found that the initial energization is due to the ion reflection and acceleration at the shock layer, the results were obtained only for one-dimensional simulations and occasionally two-dimensional simulations. As pointed out by \citet{Jokipii1993}, \citet{Giacalone1994a}, and \citet{Jones1998}, in a magnetic field that has at least one ignorable coordinate, the motions of charged particles are restricted on their original field lines of force. In this sense, the previous simulations are not conclusive. In this work, we have demonstrated that the reflection at the shock layer is the mechanism for accelerated particles gaining their initial energy in a self-consistent, three-dimensional field. However, our results are not consistent with the ``thermal leakage'' model, which assumes that the thermal particles gain the first amount of energy by thermalization process in the downstream region.

\section*{Acknowledgement}
We benefit from the discussions with Dr. Randy Jokipii and Dr. Jozsef Kota. This work was supported by NASA under grants NNX10AF24G and NNX11AO64G. Computational resources supporting this work were provided by the NASA High-End Computing (HEC) Program through the NASA Advanced Supercomputing (NAS) Division at Ames Research Center and the institutional computing resources at LANL.

\singlespace

\clearpage

\end{document}